\documentclass[prb,amsmath,amssymb,superscriptaddress,twocolumn]{revtex4}

\usepackage{graphicx}

\makeatletter
\renewcommand{\@biblabel}[1]{#1. }
\renewcommand{\@dotsep}{500}
\renewcommand{\@pnumwidth}{0em}
\renewcommand{\l@figure}[2]{
\@dottedtocline{1}{1.5em}{2em}{Figure #1}{}\vspace{15pt}}

\usepackage[normalem]{ulem}
\usepackage{upgreek}

\begin{document}

\title{Strong interactions between integrated microresonators and alkali atomic vapors: towards single-atom, single-photon operation}

\author{Roy Zektzer}\email{roy.zektzer@umd.edu}
\affiliation{Joint Quantum Institute, NIST/University of Maryland, College Park, MD 20742, USA}
\affiliation{Microsystems and Nanotechnology Division, Physical Measurement Laboratory, National Institute of Standards and Technology, Gaithersburg, MD 20899, USA}

\author{Xiyuan Lu}
\affiliation{Joint Quantum Institute, NIST/University of Maryland, College Park, MD 20742, USA}
\affiliation{Microsystems and Nanotechnology Division, Physical Measurement Laboratory, National Institute of Standards and Technology, Gaithersburg, MD 20899, USA}

\author{Khoi Tuan Hoang}
\affiliation{Joint Quantum Institute, NIST/University of Maryland, College Park, MD 20742, USA}

\author{Rahul Shrestha}
\affiliation{Joint Quantum Institute, NIST/University of Maryland, College Park, MD 20742, USA}

\author{Sharoon Austin}
\affiliation{Joint Quantum Institute, NIST/University of Maryland, College Park, MD 20742, USA}
\affiliation{Joint Center for Quantum Information and Computer Science, NIST/University of Maryland, College Park, MD 20742, USA}

\author{Feng Zhou}
\affiliation{Joint Quantum Institute, NIST/University of Maryland, College Park, MD 20742, USA}
\affiliation{Microsystems and Nanotechnology Division, Physical Measurement Laboratory, National Institute of Standards and Technology, Gaithersburg, MD 20899, USA}

\author{Ashish Chanana}
\affiliation{Microsystems and Nanotechnology Division, Physical Measurement Laboratory, National Institute of Standards and Technology, Gaithersburg, MD 20899, USA}

\author{Glenn Holland}
\affiliation{Microsystems and Nanotechnology Division, Physical Measurement Laboratory, National Institute of Standards and Technology, Gaithersburg, MD 20899, USA}

\author{Daron Westly}
\affiliation{Microsystems and Nanotechnology Division, Physical Measurement Laboratory, National Institute of Standards and Technology, Gaithersburg, MD 20899, USA}

\author{Paul Lett}
\affiliation{Joint Quantum Institute, NIST/University of Maryland, College Park, MD 20742, USA}
\affiliation{Quantum Measurement Division, National Institute of Standards and Technology, Gaithersburg, MD 20899, United States of America}

\author{Alexey V. Gorshkov}
\affiliation{Joint Quantum Institute, NIST/University of Maryland, College Park, MD 20742, USA}
\affiliation{Joint Center for Quantum Information and Computer Science, NIST/University of Maryland, College Park, MD 20742, USA}
\affiliation{Quantum Measurement Division, National Institute of Standards and Technology, Gaithersburg, MD 20899, United States of America}

\author{Kartik Srinivasan}\email{kartik.srinivasan@nist.gov}
\affiliation{Joint Quantum Institute, NIST/University of Maryland, College Park, MD 20742, USA}
\affiliation{Microsystems and Nanotechnology Division, Physical Measurement Laboratory, National Institute of Standards and Technology, Gaithersburg, MD 20899, USA}


\begin{abstract}
Cavity quantum electrodynamics (cQED), the interaction of a two-level system with a high quality factor ($Q$) cavity, is a foundational building block in different architectures for quantum computation, communication, and metrology. The strong interaction between the atom and the cavity enables single photon operation which is required for quantum gates and sources. Cold atoms, quantum dots, and color centers in crystals are amongst the systems that have shown single photon operations, but they require significant physical infrastructure. Atomic vapors, on the other hand, require limited experimental infrastructure and are hence much easier to deploy outside a laboratory, but they produce an ensemble of moving atoms that results in short interaction times involving multiple atoms, which can hamper quantum operations. A solution to this issue can be found in nanophotonic cavities, where light-matter interaction is enhanced and the volume of operation is small, so that fast single-atom, single-photon operations are enabled. In this work, we study the interaction of an atomically-clad microring resonator (ACMRR) with different-sized ensembles of Rb atoms. We demonstrate strong coupling between an ensemble of $\approx$50 atoms interacting with a high-quality factor ($Q>4\times10^5$) ACMRR, yielding a many-atom cooperativity $C \approx 5.5$. We continue to observe signatures of atom-photon interaction for a few ($<3$) atoms, for which we observe saturation at the level of one intracavity photon. Further development of our platform, which includes integrated thermo-optic heaters to enable cavity tuning and stabilization, should enable the observation of interactions between single photons and single atoms.     
\end{abstract}

\maketitle
\section{Introduction}

\noindent  Single photon gates, sources, detectors, and memories \cite{Gisin2007,Monroe2014,Uppu2021,Lei2023,Sangouard2011} are amongst the basic building blocks for optical quantum computers, simulators, and communications systems~\cite{georgescu_quantum_simulation_review,reiserer_colloquium_cQED_quantum_networks}. In most architectures, a key aspect of these devices is the ability to generate a strong interaction between single photons and a medium. Chip-scale integration of such devices has the advantage of scalability and miniaturization and the opportunity to decrease the interaction volume and improve light-matter interaction, particularly in the context of cavity quantum electrodynamics (cQED)~\cite{haroche2006exploring}. Indeed, by integrating single solid-state quantum emitters with nanophotonic cavities, single-photon sources and devices exhibiting nonlinearity at the single-photon level have been demonstrated \cite{aharonovich_solid-state_2016,janitz_diamond_cQED_review,uppu_quantum-dot-based_2021,Srinivasan2007,Dibos2018}. Alternately, single cold atoms coupled to  a nanophotonic cavity have been used for gate formation \cite{Tiecke2014,Bechler2018}, while ensembles of rare earth ions embedded in nanophotonic cavities have been used as a quantum memory \cite{Zhong2017,Craiciu2019}. Heterogeneous or hybrid integration \cite{Elshaari2020} further allows the incorporation of such devices with the established silicon nitride (Si$_3$N$_4$, hereafter SiN) photonic integrated circuit platform, as was done for on-demand single-photon sources via wafer bonding and subsequent device processing~\cite{Davanco2017} or by placing an already fabricated superconducting single photon detector directly on a waveguide~\cite{Najafi2015}. While there are thus many systems through which foundational quantum resources can be constructed, they all typically require significant physical infrastructure for operation. This includes cryostats for solid-state systems and magneto-optical trapping in ultra-high vacuum for cold atom systems.

Recently, a comparatively simple quantum technology has been demonstrated, the atomic-cladded waveguide (ACWG)~\cite{Stern2013}, following earlier works exploring the interaction of atomic vapors with hollow core fibers~\cite{Venkataraman2013,Perrella2018_hollowCore}, hollow core waveguides ~\cite{Yang2007} and tapered fibers~\cite{Spillane2008_selimTaper,Jones2015_pittmanEIT,Finkelstein2021}. The technology is based on bonding an alkali vapor cell onto a photonic chip, resulting in a standalone system that can then operate at room (or slightly elevated) temperature and does not require any vacuum, cryogenic, or trapping infrastructure. Such devices have presented chirality~\cite{Zektzer2019,Stern2018}, strong nonlinearity in the telecom regime~\cite{Zektzer2021_tel,Skljarow2020} and Autler-Townes splitting~\cite{Stern2017}. A key parameter to evaluate light-matter interaction in cQED is the single-atom cooperativity, $C_{0}=g_{0}^2/2\kappa\gamma$, where $g_0$ is the coupling between the atom and the cavity and $\kappa$ and $\gamma$ are the cavity field decay and atomic dephasing rate, respectively. The short-time interaction of fast-moving atoms (300~m/s) with a nanoscale optical mode limits the atomic state’s coherence~\cite{Jones2015_pittmanEIT}, resulting in $\gamma/2\pi\approx200$~MHz~\cite{Stern2013,Alaeian2020} (instead of the natural linewidth of 6 MHz). The excess broadening can affect metrological applications \cite{Zektzer2021_nat_ph}, and may also impact single-photon operations, at a minimum, by limiting the number of operations that can occur before the atom leaves the cavity. The transit time limitation can be avoided for metrological applications by diffracting light from the waveguide into a large, free-space beam that is emitted from the photonic chip~\cite{Hummon2018,Sebbag2021}, but this will reduce light-matter interaction significantly. On the other hand, it has been suggested that the large coupling strengths in nanophotonic cavities can overcome the excess broadening in warm atoms resulting from transit time, so that even single-atom-single-photon strong coupling in cQED may be observable~\cite{Alaeian2020}. Working towards that regime, collective cooperativity of $\approx~$42 has recently been demonstrated with an 80~$\mu$m radius microdisk resonator coupled to a dense ensemble of Rb atoms~\cite{Naiman2021}, following earlier work in which Rb interactions with low quality factor cavities were studied~\cite{ritter_coupling_2016,stern_enhanced_2016}.

Here, we develop an integrated photonics platform (Fig.~\ref{Fig1}(a)) that combines air-clad microring resonators (ACMMRs) and Rb vapor, with integrated buried heaters enabling a resonator mode to be frequency locked for long-term stable operation. We demonstrate strong coupling between an ensemble of $\approx100$ atoms interacting with a high-Q ($>4\times10^5$, $\kappa/2\pi\approx445$~MHz) and small volume (20 $\mu$m radius) cavity mode, with a many-atom coupling strength $g/2\pi\approx1$ GHz and many-atom cooperativity $C\approx5.5$ achieved ($C=g^2/2\kappa\gamma$, and $g=\sqrt{N}g_{0}$ and $C=NC_0$, where N is the number of atoms, if all atoms interact with the field uniformly). We also study these vapor cQED devices at lower atomic densities, and while vacuum Rabi splitting is no longer seen at the lowest density, we continue to observe saturation effects in the cavity transmission when <$3$ atoms are in the cavity on average, with saturation occuring for $\approx1$ photon in the cavity on average. With improvements to the cavity parameters -- in particular a reduction in size to increase $g_{0}$, for example using photonic crystal ring resonator devices~\cite{Lu2022_phc,Lu_rod} -- strong coupling between single vapor-phase atoms and single photons should be achievable.

\section{Initial Results and Improved Device Platform}

\begin{figure}[!t]
\centering\includegraphics[width=\linewidth]{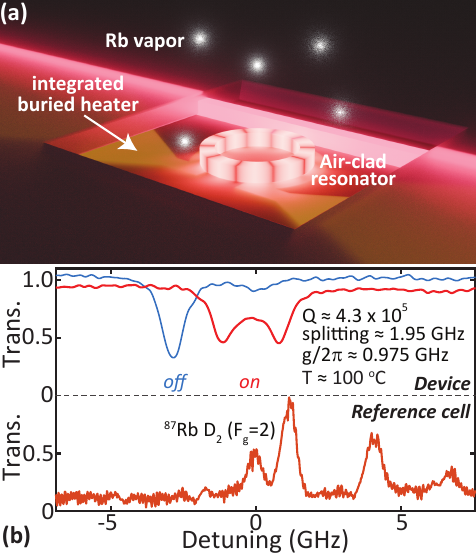}
\caption{ \textbf{System and initial results.} \textbf{(a)} Illustration of the platform developed in this work, where atoms from a Rb vapor interact with the evanescent field of an air-clad resonator. We utilize integrated heaters that are buried underneath the lower waveguide cladding for resonator tuning and stabilization, enabling measurements in the low-atom, low-photon regime.  \textbf{(b)} Initial results in a device without buried heaters. (Top) Representative cavity transmission measurement off-resonance (blue) and on-resonance (red) with the $^{87}$Rb $D_2$ transition ($5S_{1/2} (f=2)\rightarrow 5P_{3/2} (F=1,2,3)$) of an ensemble of $\approx$50 Rb atoms (temperature of 100~$^{\circ}$C). These measurements are taken from a device without integrated buried heaters; measurements in subsequent figures use stabilization. (Bottom) Simultaneously recorded absorption spectrum of a Rb reference cell containing both $^\text{87}$Rb and $^\text{85}$Rb (the cavity device only contains $^\text{87}$Rb).}
\label{Fig1}
\end{figure}

Our device platform concept is illustrated in Fig.~\ref{Fig1}(a), where moving atoms interact with the micro ring resonator (MRR) mode and integrated heaters enable control of the cavity mode frequency. Before implementing this full platform, we first consider devices without the heaters and focus on understanding the potential degradation of cavity $Q$ in the presence of Rb vapor. We fabricate SiN MRR devices with a radius of 20~$\mu$m, thickness of 250~nm, and ring width of 1~$\mu$m according to a standard SiN process flow, bond a borosilicate glass cell to the chips using vacuum epoxy, and send them to a commercial vendor who evacuates and fills them with pure $^{87}$Rb (see Appendix). As the chip facets are outside the cell region, coupling into and out of the chips can be done using standard approaches with lensed optical fibers. By keeping the cell pinch-off point sufficiently far from the photonic chip, we can limit the potential contamination of the integrated photonics devices during the cell filling process.

We install the devices in a measurement setup that has dedicated heater mounts for both the chip and the cell (see Appendix). Setting each at a temperature of 100~$^{\circ}$C, we first perform swept-wavelength spectroscopy of cavity modes that are off-resonance from the Rb transitions (top panel of Fig.~\ref{Fig1}(b)), along with simultaneous characterization of a reference Rb cell containing both $^{87}$Rb and $^{85}$Rb (bottom panel of Fig.~\ref{Fig1}(b)). We measure optical modes with $Q$ as high as $\approx4.3{\times}10^5$ when detuned from the Rb transitions. By adjusting the chip temperature, we tune the mode into the $^{87}$Rb F=2 transition and observe its pronounced splitting, by $\approx 1.95$~GHz. We note that the $^{87}$Rb F=1 transition (the rightmost peak in the reference cell spectrum) is too far-detuned from the cavity mode to interact with it, while unlike the reference cell, the device does not contain $^{85}$Rb (the two central peaks in the reference cell spectrum). Thus, a simple single splitting of the cavity mode is observed.

The splitting in the on-resonance spectrum is twice the many-atom coupling rate $g/2\pi\approx1$~GHz. The cavity field decay rate ($\kappa$) is given by the off-resonance half linewidth ($\approx445$~MHz). Assuming a decay rate $\gamma/2\pi\approx200$~MHz (controlled by the transit time), we obtain a collective cooperativity $C=g^2/2\kappa\gamma\approx$~5.5. From the cell temperature of 100~$^{\circ}$C, we expect an equilibrium Rb density of 4.7$\times$10$^{18}$ cm$^{-3}$~\cite{alcock1984vapour}, which for a simulated cavity volume of 11.2~$\mu$m$^{-3}$, gives an estimate of $N_\text{at}\approx$53 atoms interacting with the cavity. The average single-atom cooperativity $C_0=C/N_\text{at}\approx0.1$, which assumes that every atom interacts with the field equally. As discussed in the Appendix, many-atom cavity QED simulations that take into account the spatial variation of the field and which average over many distributions of the atomic velocities and spatial positions are consistent with an atom in the strongest part of the cavity field experiencing $g_{0}/2\pi\approx330$~MHz, corresponding to $C_0\approx0.61$.

\begin{figure}[!t]
\centering\includegraphics[width=\linewidth]{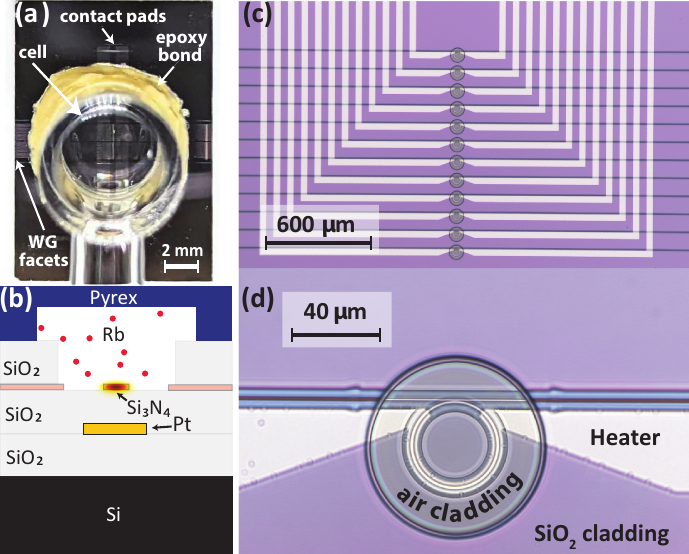}
\caption{ \textbf{Experimental platform: an atomically-clad microring resonator (ACMRR) with integrated buried heaters.} \textbf{(a)} Photograph of a completed device, with important features indicated. \textbf{(b)} Cross-sectional schematic of the vapor-cavity interaction region. \textbf{(c)-(d)} Optical microscope images of the fabricated photonic chip, with the heater traces, SiO$_2$ cladding, and air cladding regions indicated.}
\label{Fig2}
\end{figure}

These initial measurements verify that high $Q$ can be preserved in the presence of Rb and that vacuum Rabi splitting due to the contribution of a few dozen atoms can be observed. To probe this system more carefully, and in particular in the limit of a small number of atoms and with a single-photon-level intracavity field requires long averaging times, which demands fine control over the cavity detuning and elimination of any thermal drifts, as 10~mK shifts in temperature result in $\approx$100~MHz shifts in the cavity mode frequency. Next, we will explain how our new platform and experimental setup achieve this needed stability. The platform (Fig.~\ref{Fig2}(a)-(d)) consists of a 250~nm thick Si$_3$N$_4$ waveguide layer, underlying metallic heater and lower SiO$_2$ cladding layers, and an upper cladding that is air near the resonators (for Rb interaction) and SiO$_2$ everywhere else (for cell bonding and facet coupling), with more details provided in the Appendix. Figure~\ref{Fig2}(a) shows a photograph of a completed device, highlighting that both the waveguide facets and electrical contact pads are outside the cell region and therefore accessible. Figure~\ref{Fig2}(b) shows a schematic cross-section indicating the material layer stack, while Fig.~\ref{Fig2}(c)-(d) shows optical microscope images of a fabricated photonic chip, indicating the air-clad resonator, SiO$_2$-clad waveguide, and heater regions.

\section{Setup stability and anti-crossing spectroscopy}

\begin{figure}[t!]
\centering\includegraphics[width=\linewidth]{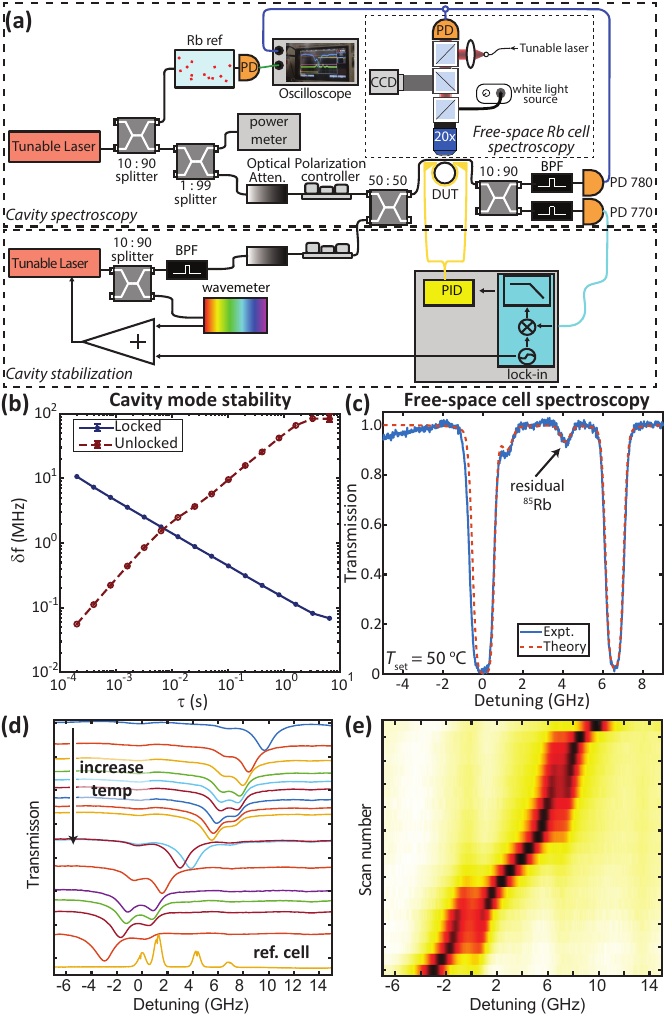}
\caption{ \textbf{Measurement setup and anti-crossing spectroscopy.} \textbf{(a)} Schematic illustration of the measurement setup. The top dashed box indicates the cavity spectroscopy portion, where a tunable laser is split to simultaneously interrogate a reference Rb cell and the cavity-vapor device at wavelengths near the Rb D2 transition. The same laser can be sent to interrogate the device in free-space (inner dashed box), to provide a consistency check on the vapor density estimated from the set cell temperature. The bottom dashed box indicates the cavity stabilization portion, based on simultaneous probing of a second cavity resonance (detuned from Rb) and feedback to chip-integrated heaters. PD: photodetector; CCD: charge-coupled device; DUT: device under test; BPF: bandpass filter; PID: proportional-integral-derivative. \textbf{(b)} Allan deviation of the cavity mode frequency shift over time when locked (blue) and unlocked (red). \textbf{(c)} Representative free-space spectroscopy of the device cell region at a temperature of 50~$^{\circ}$C, along with the expected result from a theoretical model (dashed line). \textbf{(d-e)} Device cavity transmission for different detunings of the MRR from the $^{87}$Rb D2 transitions, showing vacuum Rabi splitting and anti-crossing. The cavity temperature is changed by $\approx1~^{\circ}$C from the top scan to the bottom scan using the buried integrated heaters. The bottom spectrum in \textbf{(d)} is taken from the reference cell.}
\label{Fig3}
\end{figure}

Our measurement setup for interrogating the ACMRRs with integrated buried heaters is depicted in Fig.~\ref{Fig3}(a). We place fabricated samples in a setup in which both the chip and glass cell have independent proportional-integral-derivative (PID) temperature control (see Appendix) along with the device-level fine temperature control provided by the integrated heaters. We proceed to tune a high-$Q$ resonance to the $^{87}$Rb $F=2$ transition in a coarse fashion by setting the chip holder temperature to $\approx$~117~$^{\circ}$C. To fine-tune the position of the resonance and ensure stability for long-term measurements, we stabilize the MRR temperature by locking its mode around 770~nm to a laser using the buried heaters~\cite{Zektzer2016}, with the laser wavelength stabilized by feedback from a wavemeter. The device interrogation is done by a second laser at 780 nm. Cross-talk between the locking system and measurement system is avoided by placing bandpass filters before and after the device. Using the error signal from the lock-in amplifier we estimate the MRR mode frequency stability and extract its Allan deviation when locked and unlocked (Fig.~\ref{Fig3}(b)). The MRR long-term stability drifts by more than $\approx100$~MHz when unlocked, which is on par with the mode linewidth and as such is a problem for long-term measurements. However, when locked, the mode stability is at the $<1$~MHz level, making long-term measurements possible. 

We seek to probe the devices in the regime of small numbers of atoms and photons in the cavity. The number of photons in the cavity is controlled by the input power into the waveguide on the chip, and is on the order of 1~nW for an average intracavity photon number near 1. The number of atoms is determined by the vapor temperature, which is controlled by a heater embedded within a copper shroud that surrounds the cell (see Appendix). We estimate the vapor temperature using a free-space system (Fig.~\ref{Fig3}(a)) that interrogates the cell through its top window.  Measuring the transmission of a signal reflected from the chip and fitting it to a standard model \cite{Siddons_2008_rb_abs} (Fig.~\ref{Fig3}(c)) provides an estimate of the vapor temperature that is generally consistent with the temperature reading on the copper shroud. 

Next, we tune the cell temperature to 120~$^{\circ}$C and measure the low-power device transmission by scanning an attenuated 780~nm laser across the cavity resonance. We tune the chip temperature by $\approx$1~$^{\circ}$C using the buried heaters to shift the MRR mode across the Rb D2 transitions. We present a sample of the data in Fig.~\ref{Fig3}(d) and the full set of spectra in Fig.~\ref{Fig3}(e). Clear anti-crossing and Rabi splitting are observed as the cavity crosses both $^{87}$Rb D2 ground state transitions.

\section{Measurements at reduced atom number}

\begin{figure}[t!]
\centering\includegraphics[width=\linewidth]{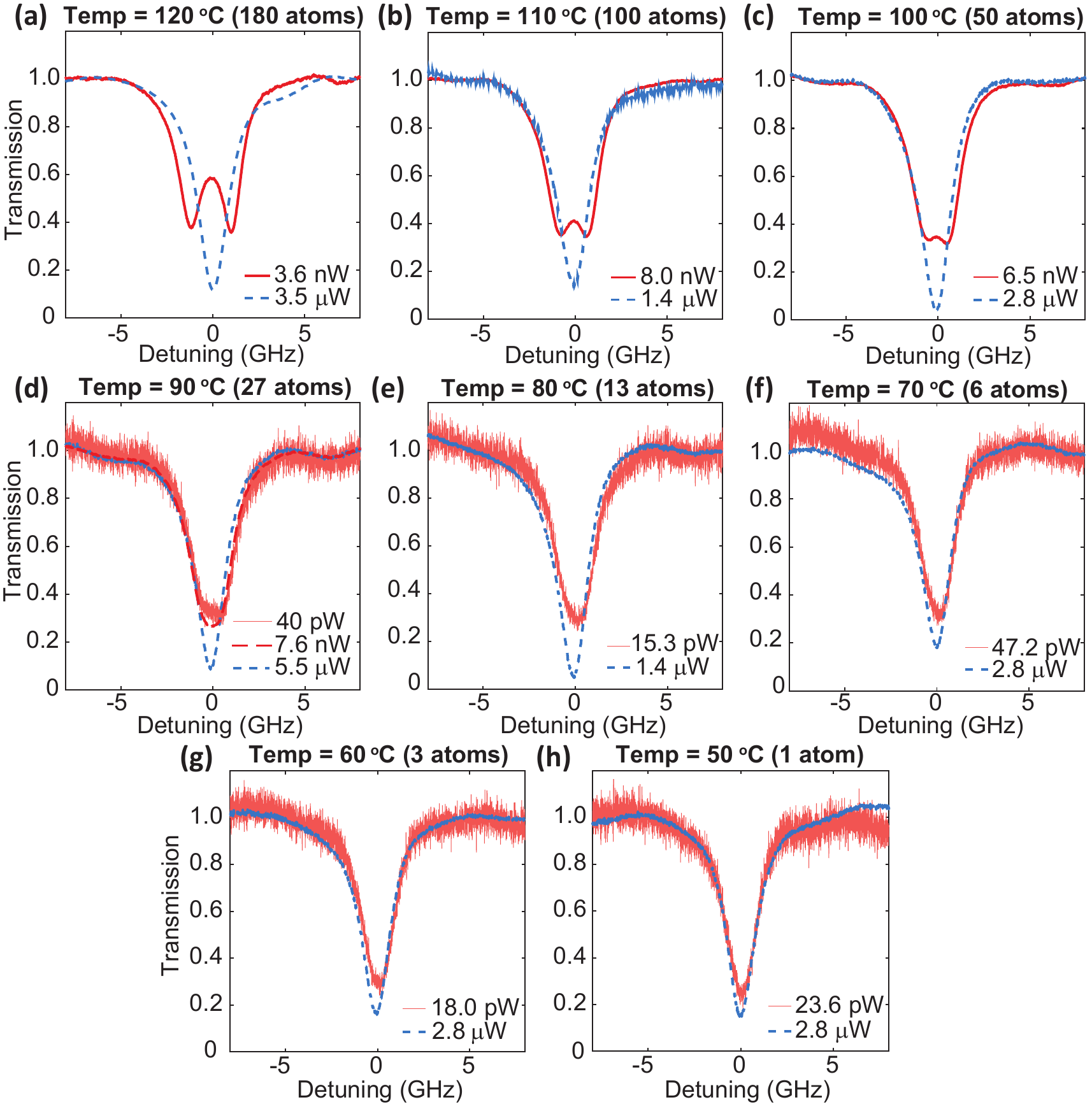}
\caption{ \textbf{High power, low power, and few-photon-level cavity transmission as a function of vapor temperature.} 
\textbf{(a)-(c)} High power (dashed blue) and low power (dark red) cavity transmission spectrum when on resonance with the Rb F=2 transition at a vapor temperature of 120~$^{\circ}$C (a), 110~$^{\circ}$C (b), and 100~$^{\circ}$C (c). \textbf{(d)} High power (dashed blue), low power (dashed dark red), and few-photon-level power (light red) at a vapor temperature of 90~$^{\circ}$C. \textbf{(e)-(h)} High power (dashed blue) and few-photon level (light red) cavity transmission spectrum when on resonance with the Rb F=2 transition at a vapor temperature of 80~$^{\circ}$C (e), 70~$^{\circ}$C (f), 60~$^{\circ}$C (g), and 50~$^{\circ}$C (h). As in \textbf{(d)} a slight difference is observed between the few-photon-level and low-power data, indicating the onset of atomic saturation, in \textbf{(e)}-\textbf{(h)} we focus on the comparison between the few-photon-level and high-power results. The estimated number of atoms within the cavity near-field is listed in each sub-panel.} 
\label{Fig4}
\end{figure}

We next study the on-resonance saturation behavior of the system, with the cavity spectral position fixed using a wavelength modulation locking setup with feedback to the buried heaters, as indicated in Fig.~\ref{Fig3}(a). We investigate the atom-cavity interaction at various cell temperatures and input powers (Fig.~\ref{Fig4}) and observe reduced splitting as we saturate the atoms (i.e., at higher optical powers) and decrease the atomic density. Splitting is observed down to $\approx 50$ atoms in the cavity, yet is slightly less prominent compared to the results shown in Fig.~\ref{Fig1} as the device with the buried heaters has a $Q\approx2{\times}10^5$ compared to $Q\approx4{\times}10^5$ of the device without heaters. We do not believe that the reduced $Q$ is due to the heaters themselves, given previous results showing $Q\approx7.5{\times}10^5$ with heaters incorporated~\cite{Moille2022}. Instead, the reduced $Q$ is potentially due to different lithography conditions when patterning the microrings. While splitting is no longer present at the lowest densities, an observable difference between the low power and high power spectrum, and hence a signature of the atom-cavity interaction, was observable even at 50~$^{\circ}$C, where it is estimated that only $\approx$~1 atom interacts with the cavity on average. In particular, we observed an $\approx$~0.15 difference in the normalized transmission at 50~$^{\circ}$C between the low power signal and the saturated (high power) signal, with the latter representing the bare cavity transmission. This difference exceeds our estimated transmission variation error, which we have measured across all powers with the cavity off-resonance, and is $\approx0.1$. 

\section{Few-photon-level saturation}

\begin{figure}[t!]
\centering\includegraphics[width=\linewidth]{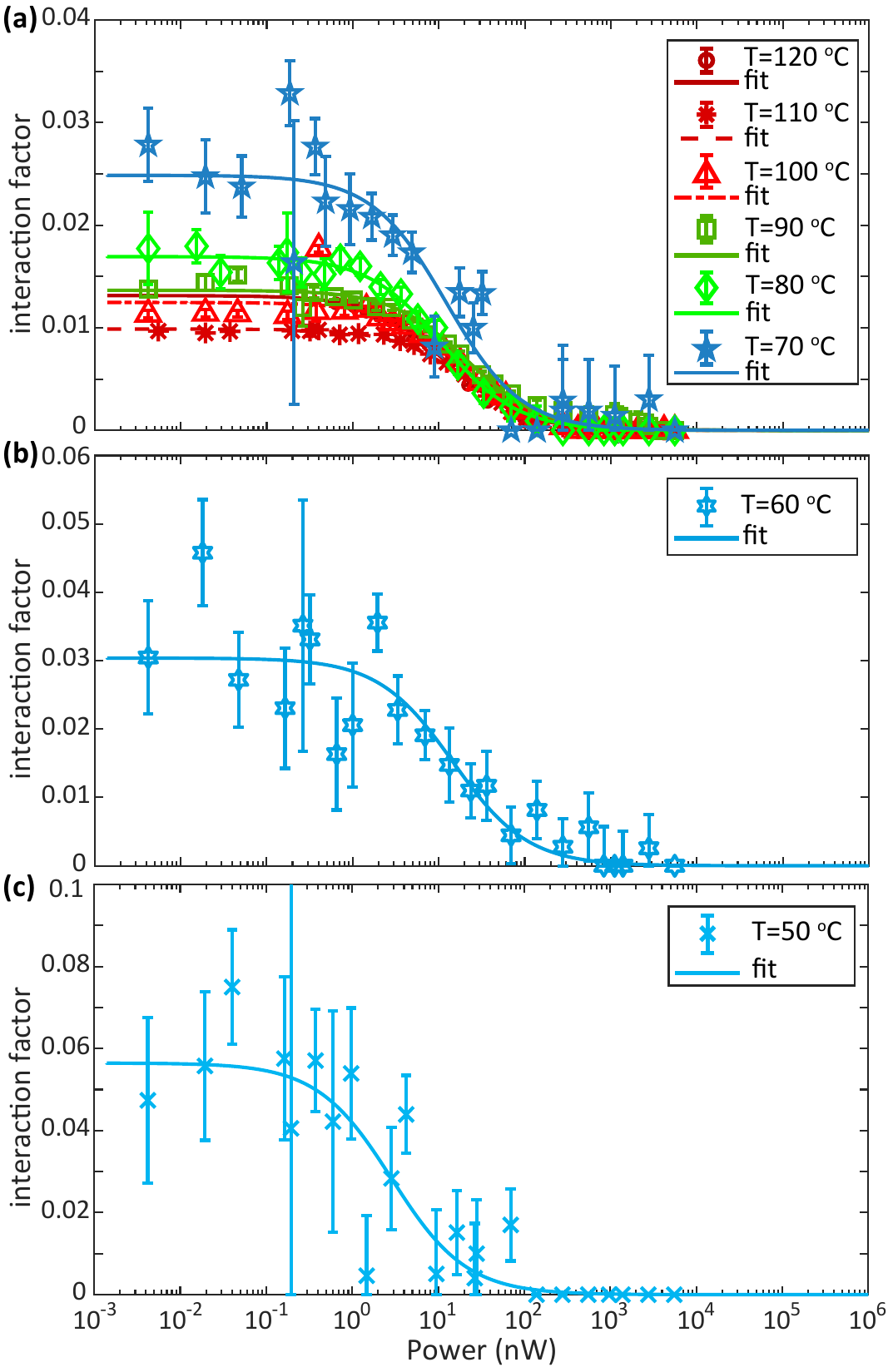}
\caption{ \textbf{Interaction factor as a function of vapor temperature.}  
\textbf{(a)} Temperatures between 120~$^{\circ}$C and 70~$^{\circ}$C. \textbf{b} Temperature of 60~$^{\circ}$C. \textbf{c} Temperature of 50~$^{\circ}$C. Symbols are experimental data points and solid curves are fits to the data. The error bars represent 95~$\%$ confidence intervals and are due to vapor temperature uncertainty and difficulty in separating fit parameters as the Rabi splitting becomes significantly smaller than the cavity linewidth.}
\label{Fig5}
\end{figure}

To better understand the saturation behavior of the system and estimate the intracavity saturation photon numbers, we fit our transmission spectra to the transfer function of an ACMRR following the Stern and Levy formalism~\cite{Stern2012} (see Appendix). We calculate the expected round-trip loss by evaluating the refractive index of Rb in the vicinity of a waveguide~\cite{Stern2013} and using it in a finite-element method simulation to extract the modal loss. The modal loss is multiplied by a free parameter to accompany the saturation effect. In Fig.~\ref{Fig5}, we present this parameter, which we term as an interaction factor, for the different measurements taken at varying cell temperatures. We then fit the data to the known expression $\alpha=\alpha_0(1/(1+P/P_{sat}))$, where $\alpha$ is the interaction factor, $\alpha_0$ is a constant prefactor, $P$ is the power, and $P_{sat}$ is the saturation power. From this expression, we have extracted $P_{sat}$ for different atomic densities. In Fig.~\ref{Fig5}(a), we display the saturation data for cell temperatures between 120~$^{\circ}$C and 70~$^{\circ}$C, and observe a clear saturation behavior in all cases. We can still see a saturation effect at 60~$^{\circ}$C and 50~$^{\circ}$C (Fig.~\ref{Fig5}(b)-(c)), though the uncertainty is greater. Across the full set of data, we observe saturation powers as high as 20~nW (higher temperature/atomic density) and as low as 3~nW (lowest temperatures/atomic density). We can deduce the number of photons in the cavity ($n_{cav}$) using $n_{cav}=\Delta T P_{in} Q_T/((1+\sqrt{1-\Delta T})\hbar \omega_0^2$), assuming an undercoupled cavity (as is the case in our system). \cite{Srinivasan2008Ncav}. Using a measured loaded quality factor ($Q_T$) of $2.2\times10^5$, transmission contrast ($\Delta T$) of 0.8, input power ($P_{in}$) of 3 nW, and angular frequency ($\omega_0$) on the Rb D2 resonance, we estimate $n_{cav}\approx0.7$. 

We can compare the estimated $n_{cav}$ from our experiments with the saturation photon number $n_{sat}$ typically used in cavity QED~\cite{Kimble1998}, given as $n_{sat}=\gamma^2/(2g_{0}^2)$, where $\gamma$ and $g_0$ are as previously defined, with $\gamma/2\pi\approx200$~MHz due to transit time broadening. As the warm, moving atoms are interacting with different parts of the cavity field in all measurements, we estimate an average $g_0$ ($\bar{g}_0$) based on the Rabi splitting data taken at high temperature (100~$^{\circ}$C). We take $\bar{g}_0=g/\sqrt{N_{at}}$, with $g$ being half the Rabi splitting for an on-resonance transmission spectrum in Fig.~\ref{Fig4}(e). As discussed earlier, we estimate $N_{at}$ by calculating the mode volume that can interact with the atoms (using the mode decay length) and multiplying it by the atomic density at this temperature (we choose 100~$^{\circ}$C). Using this approach, we deduce $\bar{g}_0/2\pi\approx125$~MHz, so that $n_{sat}\approx1.3$, on par with our measured results. 

To understand the saturation behavior as we increase the number of atoms in the cavity, we present a simple model of atoms as oscillators inside a cavity (see Appendix), which gives rise to the following expression $P_{sat}\sim(1+NC_0)^2/C_0$. For the device used in this study, at 100~$^{\circ}$C the collective cooperativity ($C=NC_0$) is estimated to be 1.69, so that $C_0\approx0.033$. Our model predicts a six-fold increase in the saturation power as we increase the number of atoms to $\approx$50, which is similar to the increase we observe. Finally, one point of interest is in comparing the ACMRR to the ACWG with regards to single-atom operation and nonlinearity. To observe a signal at such densities with the ACWG one needs to interact with millimeter long waveguides~\cite{Stern2017}, which will result in the interaction with a large ensemble of atoms. In regards to nonlinearity, saturation power at the 1~nW level has been reported for waveguide devices \cite{Stern2013,Zektzer2021_nat_ph}, and similar switching powers have been reported as well~\cite{Zektzer2021_tel,Skljarow2020,gaetafewPhoton,Spillane2008_selimTaper}. Thus, when it comes to realizing low power nonlinearities, the advantages of the ACMRR may be limited in comparison to the ACWG.  However, for single-photon, single-atom applications such as gates and single-photon sources, the ACMRR is probably preferred.

%
%
\section{Discussion}
In summary, we have fabricated an MRR with buried integrated heaters and bonded it to a small rubidium cell. We used fine temperature control to study the atom-light interaction and observed peak collective cooperativity of 5.5 and single-atom cooperativity of 0.1. We have measured our device at different atom densities and observed nonlinear interaction in the few atom regime ($\approx3$), with saturation observed for approximately 1 photon in the cavity. Such nonlinearity may already be useful for quantum applications~\cite{Pick2021,Chang2014SinglePhNonlinearity}, while collective cooperativity larger than 1 is already useful for ensemble-based memories~\cite{Afzelius2010cavityquantummemory}. The short interaction time of the atom with the cavity might limit the utility of such devices in certain applications that require long coherence times, while other applications such as photon source synchronization~\cite{Davidson2023Sync} may require fast operation. Single-atom applications such as single photon sources still require $C_0$>1, which implies the need to increase the cavity quality factor to volume ratio ($Q/V$) by a factor of 20. In principle there are several potential routes to such performance. Chip-integrated whispering gallery mode resonators have exhibited $Q\approx5\times10^8$~\cite{Kerry_integrated_highQ,blumental_highQ}, but with 1~mm scale diameters that will cause a significant reduction in $g_{0}$. Photonic crystal cavities have been used to produce a high $Q/V$ ratio~\cite{grutter_sibf_2015,HighQPhotonicCrystal,zhan_high-_2020} but they generally require much more elaborate design and simulation~\cite{LoncarPhotonicCrystal} to achieve high $Q$ compared to whispering gallery mode resonators. Photonic crystal microrings~\cite{Lu_rod,Lu2022_phc} have recently been demonstrated to preserve the high $Q$ and straightforward design of the whispering gallery mode resonators while reducing the mode volume by more than an order of magnitude, making them a compelling candidate for further investigation and optimization.


\medskip
\noindent \textbf{Funding.}
This work is supported by the DARPA SAVANT (ARO contract W911NF2120106) and NIST-on-a-chip programs. 
\\
\noindent \textbf{Acknowledgements}
We thank Archie Brown from Triad Technology for filling the Rb cells, and David Long and Peter Riley for helpful comments about the paper. Certain commercial products or names are identified to foster understanding. Such identification does not constitute
recommendation or endorsement by NIST, nor is it intended to imply that the products or names identified are necessarily the best available for the purpose.

\appendix

\section{Device Fabrication and Installation}

In this section we explain in more detail the device fabrication process and how the devices are placed and measured in the experimental setup. Device layout is done with the Nanolithography Toolbox, a free software package developed by the NIST Center for Nanoscale Science and Technology ~\cite{coimbatore_balram_nanolithography_2016}. Microring resonators are fabricated in a 250~nm thick stoichiometric silicon nitride (Si$_3$N$_4$, or SiN) layer deposited via low pressure chemical vapor deposition, through a combination of electron-beam lithograpy and reactive ion etching. After resist removal, a 3 $\mu$m SiO$_2$ layer is selectively deposited~\cite{moille_tailoring_2021} on the wafer surface through a liftoff process with a low-temperature (180~$^{\circ}$C), inductively-coupled plasma enhanced chemical vapor deposition (ICP-PECVD) process. This selective deposition ensures that the resonators remain air clad whereas the majority of the chip is oxide clad, which both limits any waveguide interaction with the Rb vapor and improves coupling to and from optical fibers. After the liftoff process is completed, chips are diced, polished, and annealed at 1100~$^{\circ}$ in an N$_2$ environment. 

\begin{figure}[ht!]
\centering\includegraphics[width=0.8\linewidth]{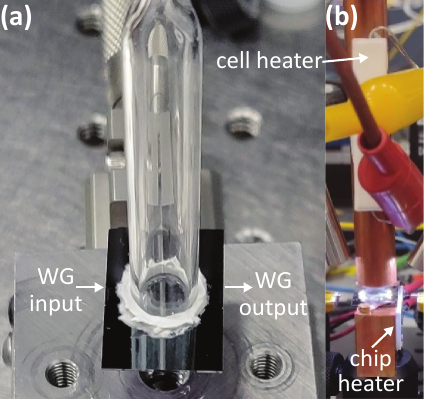}
\caption{\textbf{First generation devices.} (a) Device geometry used for the measurements in Fig.~1(b), where a straight cylindrical glass cell is epoxy-bonded to the photonic chip, leaving the polished waveguide facets accessible for fiber coupling. (b) Device installed in the measurement setup, with the chip heated through a resistive heater affixed to an underlying copper mount, and the cell heated through a resistive heater affixed to a surrounding cylindrical copper mount.}
\label{FigS1}
\end{figure}

The majority of the devices studied in this work (i.e., starting from Fig.~2 in the main text) make use of buried integrated heaters~\cite{Moille2022}. These heaters are prepared prior to the above process, through photoresist patterning and liftoff of a Cr/Pt metal layer on a 3~$\mu$m SiO$_2$ layer that is created through wet thermal oxidation of a Si substrate. 

\begin{figure*}[t!]
\centering\includegraphics[width=0.75\linewidth]{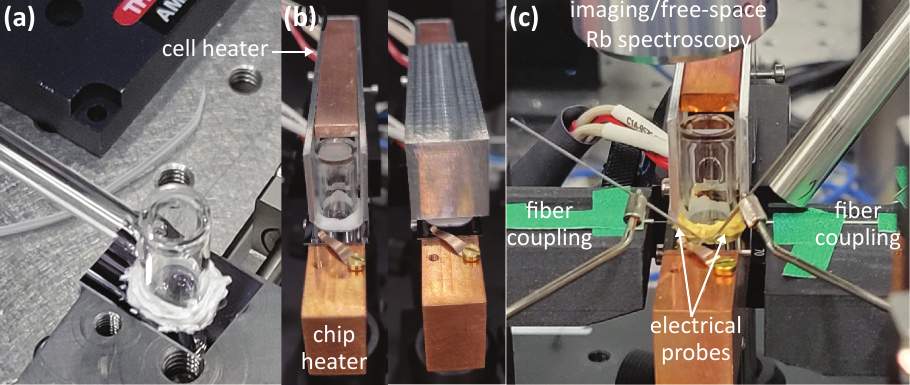}
\caption{\textbf{Second generation devices.} (a) Device geometry used for the measurements in Figs.~2-5. (b) Device installed in the measurement setup, with the chip heated through a resistive heater embedded in an underlying copper mount, and the cell heated through a resistive heater embedded in a surrounding copper and aluminum box. The box contains a lid which can be removed (left) to enable free-space access to the cell, or inserted (right) for limiting the offset between the cell temperature and box temperature. (c) View of the measurement setup including fiber coupling, electrical probes, and microscope objective that are used in characterizing the device. }
\label{FigS2}
\end{figure*}

After the metal layer definition, SiO$_2$ is deposited via low-temperature ICP-PECVD, planarized through chemical-mechanical polishing, and annealed. After the above fabrication steps, borosilicate glass cells are bonded to the chips with vacuum epoxy, with the cell diameter being sufficiently small in comparison to the chip width that the waveguide facets remain accessible. In addition, the heater traces are extended a few millimeters from the resonator region, such that the pads are located outside of the cell. The devices are then sent to a commercial vendor for filling with $^{87}$Rb, with the cells sealed using a high temperature process that melts the glass stem and forms a 'pinch-off' point that is kept far enough away from the photonic chip to limit any contamination during the filling process. Two types of cell geometries are used. For the data in Fig.~1(b) from the main text, a straight cylindrical cell oriented normal to the cell surface is used (Fig.~\ref{FigS1}(a)), and in experiments heaters attached to an underlying copper mount and cylindrical copper mount surrounding the cell are used to heat the chip and cell, respectively (Fig.~\ref{FigS1}(b)). Figure ~\ref{FigS2} shows images of the device style of relevance to the remaining data (Figs.~2-5) in the main text). Here, the cell has a top window that allows for free-space Rb spectroscopy and imaging of the microring devices, with a stem that is oriented parallel to the chip (Fig.~\ref{FigS2}(a)). Temperature is controlled through heaters embedded within the underlying sample mount and surrounding cell mount, the latter of which can be operated with a cover to ensure maximum uniformity of heating (Fig.~\ref{FigS2}(b)). A zoom-in of the experimental setup showing the orientiation of the device with respect to the fiber coupling mounts and electrical probes is shown in Fig.~\ref{FigS2}(c).


\section{Many atom cavity QED model for maximum \textbf{$g_{0}$} estimation}

One approach for modeling our system is based on a standard cavity QED formalism~\cite{carmichael2007statistical}, involving the Jaynes-Cummings Hamiltonian for multiple atoms coupled to a cavity mode, with Liouvillian terms for the cavity decay and atomic dephasing. We use a simplified model for the cavity mode, based on numerical fits to the dominant electric field component (radial polarization for a transverse electric mode). We define a simulation volume containing a number of atoms set by the atomic density at the vapor temperature (100~$^{\circ}$C) and distribute the atoms randomly within the electromagnetic field. Each atom is also assigned a transition frequency that is sampled from a Doppler distribution defined by the vapor temperature. We then average over 100 different configurations of the atoms, where in each configuration the atoms are re-distributed randomly in space and with transition frequencies that are again sampled from the Doppler distribution. 

\begin{figure}[b!]
\centering\includegraphics[width=0.75\linewidth]{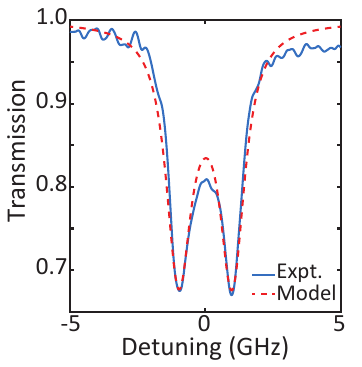}
\caption{ \textbf{Many atom cavity QED model and comparison with experiments.} Comparison at a vapor temperature of 100~$^{\circ}$C. The experimental data is taken from Fig.~1(b) in the main text.}
\label{FigS3}
\end{figure}

Figure~\ref{FigS3} shows the result of calculation compared against the experimental Rabi splitting spectrum in Fig.~1(b) from the main text. We find that the theory results are consistent with the experimental results when an atom located at the maximal point of the electric field in the air experiences a coupling strength $g_{0}/2\pi\approx$~330~MHz. This value for $g_0$ is in turn consistent with expectation based on a calculation of the effective mode volume of the resonator. That being said, the model contains many simplifications, for example, in the description of the atoms as simple two-level systems, the cavity field consisting of a single polarization, and motion of the atoms neglected other than introducing an ad hoc transit time broadening of 200~MHz in the Liouvillian term.  Nevertheless, it provides some qualitative connection between the current many-atom strong coupling experiments and potential future single-atom strong coupling experiments.  Finally, we note that this peak value of $g_0$ is about 2.5$\times$ larger than the average value $\bar{g}_0$ that is extracted directly from the measured Rabi splitting and estimated number of atoms in the cavity volume.

\section{Effective atomic cladding medium and device transfer function fits}

\begin{figure}[t!]
\centering\includegraphics[width=\linewidth]{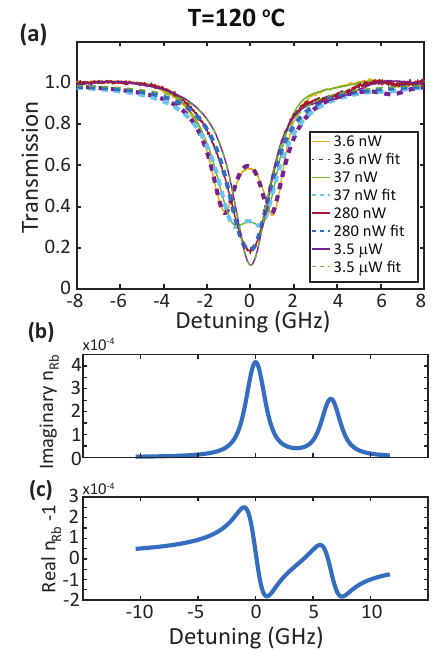}
\caption{ \textbf{Comparison between experiment and theory.} 
\textbf{(a)} Measurements (solid lines) and fits (dashed lines) at 120~$^{\circ}$C for different power levels. \textbf{(b)} Imaginary part of the Rb refractive index. \textbf{(c)} Deviation of the real part of the Rb refractive index from 1.}
\label{FigS4}
\end{figure}

We use the theoretical formalism developed by Stern and Levy~\cite{Stern2012} to fit the experimental data in Figs.~4 and 5 in the main text. First, we simulate the effective index of the mode ($n_{eff}$). Using the wavenumber of the mode in the propagation and the transverse direction we can estimate the Rb refractive index ($n_{Rb}$) from its susceptibility according to eq.~(1) in \cite{Stern2012}. The field at the output of the MRR is as follows:

\begin{equation}
E_{out}=\frac {r-\tau\cdot e^{i\cdot k\cdot L}}{1-r\cdot \tau \cdot e^{i\cdot k \cdot L}}
\end{equation}

\noindent where $r$ is the waveguide-to-ring coupling coefficient and $\tau$ is the round-trip transmission amplitude (1-loss).  The mode wavenumber ($k$) is re-evaluated using the rubidium effective index and the interaction factor ($IF$) takes into account the percentage of the mode interacting with the Rb and the saturation effect. The Rb refractive index is comprised of small variations in the real part around 1 and small variations in the imaginary part around zero (Fig.~\ref{FigS4}(b)-(c)). As we are interested in the dispersive elements of the lineshape, we consider the deviation in the real part of the refractive index from 1:

\begin{equation}
k=\frac{2\pi((n_{Rb}-1)\cdot IF+n_{eff})}{\lambda}
\end{equation}

We fit our transmission to $|E_{out}|^2$ using $IF$ as the fitting parameter. Several fit examples are shown in Fig.~\ref{FigS4}(a).

\section{Many oscillator model for saturation}

\noindent To provide a simple prediction for the onset of device saturation as we adjust the atomic density, we look at our system as a set of $N$ oscillators (atoms) separately coupled to another oscillator (the cavity). We then write the following equations for the evolution of the oscillator amplitudes:

\begin{equation}
\dot a=i\sum_{i=1}^{N} g\cdot b_j+ \sqrt{2\kappa}a_{in}-\kappa \cdot a 
\end{equation}
\begin{equation}
\dot b_j=i\cdot g\cdot a -\gamma b_j
\end{equation}
\noindent where $a$ is the cavity field amplitude, $a_{in}$ is the input field, $b_j$ is the amplitude of the atomic oscillator, $N$ is the number of atoms, $g$ is the atom-cavity coupling rate, $\gamma$ is the atom decay rate, and $\kappa$ is the cavity decay rate.

Solving these equations in steady-state results in:

\begin{equation}
\dot b_j=0=>b_j=\frac{i\cdot g}{\gamma\ }a
\end{equation}

\begin{equation}
\dot a=0=>a=\frac{\sqrt{2\kappa}}{\kappa (1+\frac{g^2N}{\kappa\gamma})}a_{in}=\frac{\sqrt{2\kappa}}{\kappa (1+C_N)}a_{in}
\end{equation}

\begin{equation}
b_j=\frac{i\cdot g}{\sqrt{\gamma\kappa }}\frac{\sqrt{2}}{\sqrt{\gamma} (1+C_N)}a_{in}=\frac{i\sqrt{2C_1}}{\sqrt{\gamma} (1+C_N)}a_{in}
\end{equation}

We define saturation as the atomic oscillation at an amplitude of 1 ($b_j=1$), which results in:

\begin{equation}
b_j=1=> |a_{inSat}|^2=\frac{\gamma}{2}\frac{(1+C_N)^2}{C_1}=\frac{(\gamma)}{2}\frac{(1+N\cdot C_1)^2}{C_1}
\end{equation}



\end{document}